\documentclass[prc,superscriptaddress
,nofootinbib,
tightenlines]{revtex4}

\usepackage{epsfig}
\usepackage{slashed}
\usepackage{amssymb}
\usepackage{color}
 \usepackage{hyperref}
\usepackage{orcidlink}

\newcommand{\beq}{\begin{equation}}
\newcommand{\eeq}{\end{equation}}
\newcommand{\beqa}{\begin{eqnarray}}
\newcommand{\eeqa}{\end{eqnarray}}

\newcommand{\ben}{\begin{displaymath}}
\newcommand{\een}{\end{displaymath}}
\newcommand{\be}{\begin{equation}}
\newcommand{\ee}{\end{equation}}
\newcommand{\bea}{\begin{eqnarray}}
\newcommand{\eea}{\end{eqnarray}}

\begin{document}

\title{Symmetry-preserving higher-derivative formulation of non-relativistic baryon chiral effective field theory}

 \author{X.-L.~Ren \orcidlink{0000-0002-5138-7415}}
\affiliation{Shandong Provincial Key Laboratory of Nuclear Science, Nuclear Energy Technology and Comprehensive Utilization, 
\& School of Nuclear Science, Energy and Power Engineering, Shandong University, 
250061 Jinan, China}
\author{A.~Deltuva \orcidlink{0000-0002-0732-7749}}
\affiliation
{Institute of Theoretical Physics and Astronomy, Vilnius University, Sauletekio al. 3, LT-10257 Vilnius, Lithuania}
\author{E.~Epelbaum \orcidlink{0000-0002-7613-0210}}
\affiliation
{Institut f\" ur Theoretische Physik II, Fakult\" at f\" ur Physik und Astronomie,
Ruhr-Universit\" at Bochum, D-44780 Bochum, Germany}
\author{A.~M.~Gasparyan \orcidlink{0000-0001-8709-4537}}
\affiliation
{Institut f\" ur Theoretische Physik II, Fakult\" at f\" ur Physik und Astronomie,
Ruhr-Universit\" at Bochum, D-44780 Bochum, Germany}
\author{J.~Gegelia \orcidlink{0000-0002-5720-9978}}
\affiliation
{Institut f\" ur Theoretische Physik II, Fakult\" at f\" ur Physik und Astronomie,
Ruhr-Universit\" at Bochum, D-44780 Bochum, Germany}  
\affiliation{Tbilisi State  University,  0186 Tbilisi,
 Georgia}

\date{8 November, 2026}
\begin{abstract}

A symmetry-preserving higher-derivative formulation of the non-relativistic baryon chiral effective field theory is introduced. 
Integral equations for the two-nucleon scattering and the deuteron structure functions are derived and numerical solutions at leading order are obtained.

\end{abstract}

\maketitle

\section{Introduction}

Chiral effective field theory (ChEFT) of the strong
interaction \cite{Weinberg:1978kz,Weinberg:rz,Gasser:1983yg,Scherer:2012xha,Meissner:2022cbi} has established itself as a systematic approach providing with a reliable order-by-order calculation of
physical quantities in terms of small masses and energies divided by large scales (see
e.g. Refs.~\cite{Bedaque:2002mn,Bernard:2007zu,Epelbaum:2008ga,Epelbaum:2012vx}
for reviews).  Taking symmetries of quantum chromodynamics (QCD) as input,
ChEFT aims at reproducing the $S$-matrix of QCD
in the low-energy region. Ward identities following from the underlying symmetries of QCD play an important role 
as they relate various quantities obtained from ChEFT to each other. As in any other quantum field theory, loop diagrams of ChEFT lead to ultraviolet (UV) divergences and 
therefore regularization has to be introduced. Dimensional regularization is a very convenient tool in ChEFT as it respects all symmetries of this theory. However, it cannot be applied to integral equations of the few-body sector.  
This problem is usually solved by applying cutoff regularization (in its various realizations), which violates the underlying symmetries and thus leads to complications which often become un-transparent due to solving equations numerically. 
A solution of this problem by regularizing the pion field in the effective chiral Lagrangian using the gradient flow method has been introduced recently in Ref.~\cite{Krebs:2023gge}. The actual derivation of the nuclear forces and currents from the regularized effective Lagrangian can be carried out utilizing the novel path-integral approach of Ref.~\cite{Krebs:2023ljo}.

An alternative approach to improving the ultraviolet behavior while preserving the underlying symmetries is to exploit the freedom in the formulation of the effective chiral Lagrangian, as discussed below.
To construct effective chiral Lagrangian, one has to include
{\it all} local terms which are compatible with
the underlying symmetries of QCD. However, by using various identities and field redefinitions, one usually reduces the effective Lagrangian to a (not uniquely defined) minimal form.  
Although convenient, this last step is by no means necessary. One
can equally well perform calculations using the original most general
effective Lagrangian, or any other ``non-minimal'' form. In the
context of low-energy ChEFT, this freedom can be exploited to introduce
a symmetry-preserving higher-derivative formulation 
\cite{Slavnov:1971aw,Djukanovic:2004px}, leading to integral equations of the few-body sector with improved UV behaviour.  
In current paper we work out a higher-derivative formulation, analogous to that of Refs.~\cite{Djukanovic:2006mc,Behrendt:2016nql}, applied to non-relativistic ChEFT. 
In the few-nucleon sector, we employ time-ordered perturbation theory (TOPT) and obtain integral equations satisfied by the two-nucleon scattering amplitudes and by the deuteron structure functions.

Our paper is organized as follows: In section \ref{SPF} we introduce the higher-derivative formulation of 
 the non-relativistic baryon ChEFT, section \ref{applLO} considers application to the NN scattering problem. 
In section \ref{deuteron_equation} we obtain integral equations satisfied by the deuteron structure functions,
and we summarize in section \ref{sum}.

\section{Symmetry-preserving higher-derivative formulation of baryon ChEFT}
\label{SPF}

We start with taking non-minimal effective Lagrangian of non-relativistic nucleon fields interacting with pions, where in addition to the standard canonical Lagrangian of the baryonic ChEFT (see, e.g, Ref.~\cite{Weinberg:1978kz} for the non-relativistic Lagrangian of the two-nucleon sector) we include terms with higher-order derivatives acting on the pion and nucleon fields:
\begin{eqnarray}
{\cal L}  &= & \frac{F^2}{4}\mbox{Tr}\left[D_\alpha U \left(
D^\alpha U\right)^\dagger  \right] +\frac{F^2}{4}\mbox{Tr} \left( \chi
U^\dagger+ U\chi^\dagger \right) 
+  \sum\limits_{n=1}^{\infty}
\frac{X_n}{4} \ \frac{ F^2}{4} \ {\rm Tr} \Biggl( \left\{
\left(D^2\right)^nU U^\dagger -U\left[
\left(D^2\right)^nU\right]^\dagger \right\}  \nonumber\\
&\times& \left[  D^2U
U^\dagger-U\left(D^2 U\right)^\dagger -\chi U^\dagger
+U\chi^\dagger \right] \Biggr) 
+   N ^\dagger\left( i D_0 +\frac{\vec D^2}{2 m} \right)
\left[1 + \sum_{n=1}^{\infty} Y_n \left(i D_0 +\frac{\vec D^2}{2 m}  \right)^n \right] N \nonumber\\
&+& \frac{1}{2} \, g_A N ^\dagger \sigma_i  u^i N -\frac{1}{2}\, C_S \left(N^\dagger N\right)^2 -\frac{1}{2}\, C_T \left(N^\dagger \vec \sigma N\right)^2 +\cdots \,.
\label{EFTLagr}
\end{eqnarray}
Here $X_n$ and $Y_n$ are free parameters and periods stand for higher-order terms,
the unimodular unitary $(2\times 2)$ matrix $U$ contains
the Goldstone boson fields. The covariant derivative acting on this matrix is defined as
$
D_\nu U=\partial_\nu U-i r_\nu U+i U l_\nu,
$
where
$r_\nu=v_\nu+a_\nu,$  $l_\nu=v_\nu-a_\nu,$  and $\chi =2 B (s+i p).$
The external vector,
axial-vector, scalar, and pseudo-scalar sources are represented by $v_\nu$, $a_\nu$, $s$, and $p$, respectively, and $F$ denotes the pion-decay constant in the chiral
limit.
The covariant derivative acting on the nucleon field is given by $D_\nu N = (\partial_\nu +\Gamma_\nu-i v^{(s)}_\nu)N $, where $v_\nu^{(s)}$ is the isosinglet vector field. The physical nucleon mass and the
axial-vector coupling constant in the chiral limit are given by $m$ and $g_A$, respectively. Further,
$u_\nu =iu^{\dagger} D_\nu U u^{\dagger},$
$\Gamma_{\nu}=\frac{1}{2}\left[u^\dagger\partial_{\nu}u
+u\partial_{\nu}u^\dagger- i(u^{\dagger}r_{\nu}u+u
l_{\nu}u^{\dagger})\right],$ and 
$u^2=U\,.$ 
The effective Lagrangian of Eq.~(\ref{EFTLagr}) is consistent with all underlying symmetries of ChEFT and, compared to the standard canonical form, it includes an infinite number of additional terms with higher-order derivatives. We apply the standard Weinberg power counting based on naive dimensional analysis, keeping in mind that some of the interactions might need to be promoted to lower orders. 
While the higher-derivative terms as well as the form of the canonical effective Lagrangian is fixed once, for increasing chiral orders of accuracy we include increasing numbers of higher-derivative terms in our calculations. 
These terms contribute to the nucleon two-point function as well as to interactions with pions. Contributions to the propagator are taken into account non-perturbatively, while interactions are included perturbatively according to their chiral orders. Higher-derivative terms taken into account at given order are chosen such that the integral equations for the few-body scattering amplitudes are free of UV divergences, provided that the corresponding effective potentials/kernels are already renormalized.

To regularize all ultraviolet divergences appearing in zero and one-nucleon sectors as well as in diagrams contributing to effective potentials/kernels of the few-body sector, we apply the standard dimensional regularization. Notice here that delta-function type contributions in the measure of the path integral originating from higher-derivative terms vanish within dimensional regularization \cite{Gasser:1983yg}. We express the bare field in terms of renomalized ones and fix the counter terms by choosing on-mass schell renormalization scheme.

\medskip

In general, for a specified finite order $n_\pi$ of higher-derivative pionic terms, we express the free coefficients $X_n$ in terms of other parameters $\mu_{\pi n}$ ($n=1,2,\cdots, n_\pi$) such that the resulting pion propagator is given by \cite{Djukanovic:2004px}
\begin{equation}
\Delta(p; \mu_{\pi j}) = \frac{1}{p^2-M^2+i \epsilon } \
\prod\limits_{j=1}^{n_\pi} \frac{\mu_{\pi j}^2}{\mu_{\pi
j}^2-p^2-i \epsilon } \,.
\label{genregpprop}
\end{equation}
	This propagator can be represented as a linear combination of $n_\pi+1$ terms with simple poles at $p^2=M^2$ and at $p^2=\mu_{\pi
j}^2$ ($j=1,..., n_\pi $). For $n_\pi>1$  the expressions of Eq.~(\ref{genregpprop}) can be rewritten as
\begin{equation}
\Delta(p; \mu_{\pi j})  =  \frac{1}{
p^2 - M^2 +i\epsilon } \
\prod\limits_{i=1}^{n_\pi} \frac{ - \mu_{\pi i
}^2}{M^2 - \mu_{\pi i}^2}+\sum_{j=1}^{n_\pi} \frac{ 1}{
p^2 - \mu_{\pi j}^2 +i\epsilon } \frac{- \mu_{\pi j}^2}{ \mu_{\pi j}^2-M^2}
\prod\limits_{i=1(i\neq j)}^{n_\pi} \frac{ - \mu_{\pi i
}^2}{\mu_{\pi j}^2 - \mu_{\pi i}^2} \,, \label{regppropLC}
\end{equation}
and for $n_\pi=1$ we have
\begin{equation}
\Delta(p; \mu_{\pi j}) =  \frac{1}{
p^2 - M^2 +i \epsilon} \
 \frac{ - \mu_{\pi 1
}^2}{M^2 - \mu_{\pi 1
}^2}
 + \frac{ 1}{
p^2 - \mu_{\pi 1}^2 +i \epsilon} \frac{- \mu_{\pi 1}^2}{ \mu_{\pi 1}^2-M^2} \,. \label{regppropLC1}
\end{equation}
Each term in this linear combination can be interpreted as a propagator $1/
(p^2 - M_i^2 +i\epsilon )$ ($M_i^2\in \{ M^2, \mu_{\pi
i}^2\}$) of a bosonic particle, multiplied by some coefficient. 
 
Analogously to the above case, for a specified finite order $n_N$ for the one-nucleon propagator we choose the parameters $Y_n$  in terms of other parameters $\mu_{n}$ ($n=1,2,\cdots, n_N$) such that for $n_N>1$, it takes the form
\begin{equation}
S(k;\mu_j)  = \frac{i}{k_0-\frac{\vec k^2}{2 m} +i \epsilon} \prod\limits_{i=1}^{n_N} \frac{-\mu_i}{k_0-\frac{\vec k^2}{2 m} - \mu_i +i \epsilon} 
\equiv \frac{i}{k_0-\frac{\vec k^2}{2 m} +i \epsilon} - \sum_{j=1}^{n_N} \frac{i}{k_0-\frac{\vec k^2}{2 m} -\mu_j +i \epsilon} 
\prod\limits_{i=1(i\neq j)}^{n_N} 
\frac{\mu_i}{\mu_i -\mu_j} 
\,.
\label{defReg1PPr0}
\end{equation}
Each term on the right-hand side of Eq.~(\ref{defReg1PPr0}) can be interpreted as a propagator $ i /(p_0 - p^2/2m - \delta E_i +i \epsilon  )$ ($ \delta E \in \{ 0, \mu_i \}$) multiplied by some factor. By substituting the pion and nucleon propagators in the form of linear combinations of terms with simple poles in expressions of Feynman diagrams and expanding term-by-term, each Feynman diagram can be represented as a sum of a multitude of new Feynman diagrams with new rules with fictitious particles.
Propagators of fictitious bosonic particles are given by $i/
(p^2 - M_i^2 +i\epsilon )$ ($M_i^2\in \{ M^2, \mu_{\pi
i}^2\}$), and by $ i /(p_0 - p^2/2m - \delta E_i +i \epsilon  )$ ($ \delta E_i \in \{ 0, \mu_i \}$) for fictitious fermions.  Rules for the vertices in new diagrams are generated by interaction terms of the effective Lagrangian in a somewhat complicated but systematic way. An interaction term with $i$ pion and $j$ nucleon fields generates $(n_\pi+1)^i \times (n_N+1)^j$ vertices which are obtained from the standard vertex by replacing  one-by-one the pion and the nucleon lines by lines corresponding to fictitious particles. Mathematical expressions corresponding to new vertices with differing particle contents are obtained from the original vertex by multiplying it with additional factors appearing as coefficients in linear combinations of propagators of fictitious particles as given in Eqs.~(\ref{regppropLC}) and (\ref{defReg1PPr0}). 
Notice that the inclusion of the new interaction terms, generated by additional higher-derivative terms of the effective Lagrangian, is essential for preserving the symmetries of the theory. 

All higher-derivative terms included in the Lagrangian of Eq.~(\ref{EFTLagr}) can be eliminated by field redefinitions, including their effect in the renormalization of the parameters of the canonical form. This makes explicit that $\mu_i(\mu_{\pi i})$ parameterize the choice of the renormalization condition.

\section{NN scattering}
\label{applLO}

By integrating Feynman diagrams over zeroth components of the loop momenta we obtain diagrammatic rules of TOPT for fictitious particles.
For an $N$-nucleon scattering process, we define the sum of $N$-particle-irreducible diagrams as the effective potential and obtain the scattering amplitudes by solving the corresponding integral equations. 

\medskip

To calculate the NN scattering amplitude we consider fictitious coupled-channel problem $F_I F_J \to F_K F_L$, where $I$, $J$, $K$ and $L$ run from 1 to $n_N+1$ and enumerate the  fictitious $F_1$, $F_2$, \ldots , $F_{n_N+1}$ fermions, index 1 corresponding to the fictitious particle with the standard nucleon propagator. We define the potential $V^{KL,IJ}$ of the $F_I F_J \to F_K F_L$ channel as the sum of all two-fermion-irreducible TOPT diagrams contributing to a given channel.  
 The scattering amplitudes $T^{KL,IJ}$ are obtained by solving the coupled channel equations 
\begin{equation}
T^{KL,IJ}\left({\vec p\,'} ,{\vec p}\right)=V^{KL,IJ} \left({\vec p\,'} ,{\vec p}\right) + \sum_{A,B=1}^{n_N+1} \!\int\! d^3{\vec k}\, V^{KL,AB} ({\vec p\,'} ,{\vec k} )G^{AB}({\vec k}) T^{AB,IJ} ({\vec k} ,{\vec p} \, ) \,,
\label{CChEqs}
\end{equation} 
where $G^{AB}$ is the Green's function corresponding to an intermediate $F_AF_B$ two-fermion state. 
The physical NN scattering amplitude is given by  $T^{11,11}$ calculated for the on-shell kinematics. 

\medskip

While systematic order-by-order applications are the subject of forthcoming publications, below we consider NN scattering at leading order (LO), as a simple example, for which we take $n_\pi=0$, $n_N=1$ with $Y_1=1/\mu$, and obtain the pion and nucleon propagators of the following form:
\begin{eqnarray}
\Delta(p; \mu_{\pi j})  &=&  \frac{1}{
p^2 - M^2 +i\epsilon }\,, \nonumber\\
S(k;\mu_j) &=& \frac{i}{k_0-\frac{\vec k^2}{2 m} +i \epsilon} \frac{- \mu}{k_0-\frac{\vec k^2}{2 m} - \mu  +i \epsilon} 
\equiv \frac{i}{k_0-\frac{\vec k^2}{2 m} +i \epsilon} - \frac{i}{k_0-\frac{\vec k^2}{2 m} - \mu +i \epsilon} \,.
\label{PrsLO}
\end{eqnarray}
We interpret the second term on the right-hand side of Eq.~(\ref{PrsLO}) as a propagator of fictitious particle as described in previous section. We enumerate the particles with propagators ${i}/({k_0-\frac{\vec k^2}{2 m} +i \epsilon} )$ and $ {i}/({k_0-\frac{\vec k^2}{2 m} - \mu +i \epsilon} ) $, by 1 and 2, respectively. 
We assign the factor $-1=i^2$ of the second propagator to interaction vertices. In particular, for particle 1, all interactions are taken as generated by the Lagrangian, while interactions involving particle 2 get a factor $i$ for each leg with index 2.  
By integrating over zeroth components of the loop momenta we obtain TOPT rules for our formulation in terms of fictitious particles.
In particular, particle 1 with momentum $\vec k$ has energy $E_1(k)=\vec k^2/(2m)$ and  the energy of the particle 2 with momentum $\vec k$ is $E_2(k)=\vec k^2/(2m)+\mu$\,.

\medskip

\begin{figure}[t]
\begin{center}
\epsfig{file=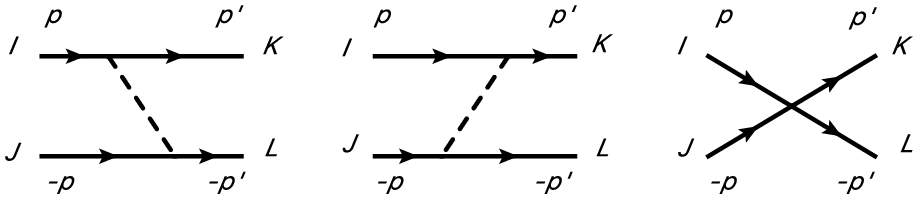,scale=0.6}
\caption{TOPT diagrams contributing to LO NN potential. Dashed lines correspond to pions and the solid lines to fermions. Here, $p$ and $p'$ correspond to three-momenta of incoming and outgoing particles, respectively.}
\label{fig1}
\end{center}
\vspace{-5mm}
\end{figure}

Diagrams contributing to the LO potential are shown in Fig.~\ref{fig1}. The corresponding expression is readily obtained by applying TOPT rules:
\begin{eqnarray}
V_{LO}^{KL,IJ}(\vec p\,',\vec p)  &=& s_I s_J s_K s_L \left( C_S + C_T  \vec \sigma_1\cdot \vec \sigma_2\right)+ s_I s_J s_K s_L \, \frac{g_A^2  \, \vec \sigma_1 \cdot (\vec p\, '-\vec p\,) \, \vec \sigma_2 \cdot (\vec p\, '-\vec p\, )}{4 F^2}\, \vec \tau_1\cdot \vec \tau_2\,\frac{1}{2\, 
\omega({\vec p\, '-\vec p}\,) } \nonumber\\
&\times& \left( \frac{1}{2 E_1(q) - E_I(p)-E_L(p') -  \omega({\vec p\, '-\vec p}\,) }+\frac{1}{2 E_1(q) - E_J(p)-E_K(p') 
- \omega({\vec p\, '-\vec p}\,)}\right) \,, 
\label{OPETOPTNSt}
\end{eqnarray}
where $ \omega(\vec k\, )  = \sqrt{\vec k^2+M^2}$, $\vec p$ and $\vec p\,’$ are the three-momenta of incoming and outgoing fermions, respectively, $E_1(q)$ is the energy of an on-shell external nucleon in the center-of-mass frame, and $s_1=1$, $s_2=i$.

The two-fermion intermediate states lead to the following two-point Green functions:
 \begin{eqnarray}
G^{11}(\vec k ) &=& \frac{m}{(2 \pi)^3} \frac{1}{m E-\vec k^2+i\epsilon} \,, \nonumber\\
G^{12}(\vec k ) &=& G^{21}(\vec k ) = \frac{m}{(2 \pi)^3} \frac{1}{
m E-\vec k^2-m \mu +i\epsilon } \,,\nonumber\\
G^{22}(\vec k )  &=& \frac{m}{(2 \pi)^3} \frac{1}{
m E-\vec k^2-2 m \mu+i\epsilon } \,,
\label{GRegs}
\end{eqnarray}
where $E=2 E_1(q)$.

By taking $\mu\leq \Lambda_b$, where $\Lambda_b$ refers to the breakdown scale of chiral EFT, loop diagrams generated by iterating the LO potential are of zeroth order, as assigned by Weinberg's power counting.

\medskip

Notice that for the pionless theory, i.e. the contact-interaction potential only, the coupled channel equations reduce to an ordinary Lippmann-Schwinger equation for the NN scattering with the two-nucleon propagator in the center-of-mass frame given by 
 \begin{equation}
G(\vec k ) = \frac{m}{(2 \pi)^3} \frac{1}{m E-\vec k^2+i\epsilon}  \frac{2 m^2 \mu^2}{
\left( m E-\vec k^2-2 m \mu+i\epsilon\right)
\left( m E-\vec k^2- m \mu +i\epsilon\right) }\,.
\label{GReg}
\end{equation}

To demonstrate that the suggested framework can lead to an adequate description of experimental data we carried out numerical calculations of the neutron-proton phase shifts at leading order for $\mu=350, \, 400, \, 500, \, 600$ MeV. 

For each value of $\mu$, the triplet contact combination
$(C_S+C_T)$ was determined by minimizing the error-weighted
$\chi^2$ of the ${}^3S_1$ phase shifts obtained from the full coupled
${}^3S_1$--${}^3D_1$ solution relative to the Nijmegen partial-wave
analysis~\cite{Stoks:1993tb} at
$T_{\rm lab}=\{1,5,10,25,50\}\,\mathrm{MeV}$.
The ${}^3D_1$ phase shifts, mixing angle $\epsilon_1$, and deuteron
binding energy are predictions. For $^1S_0$ and $^3P_0$ partial waves the corresponding coupling constants of the contact interactions have been fixed analogously.

\begin{figure}
\includegraphics[width=0.95\textwidth,keepaspectratio,angle=0,clip]{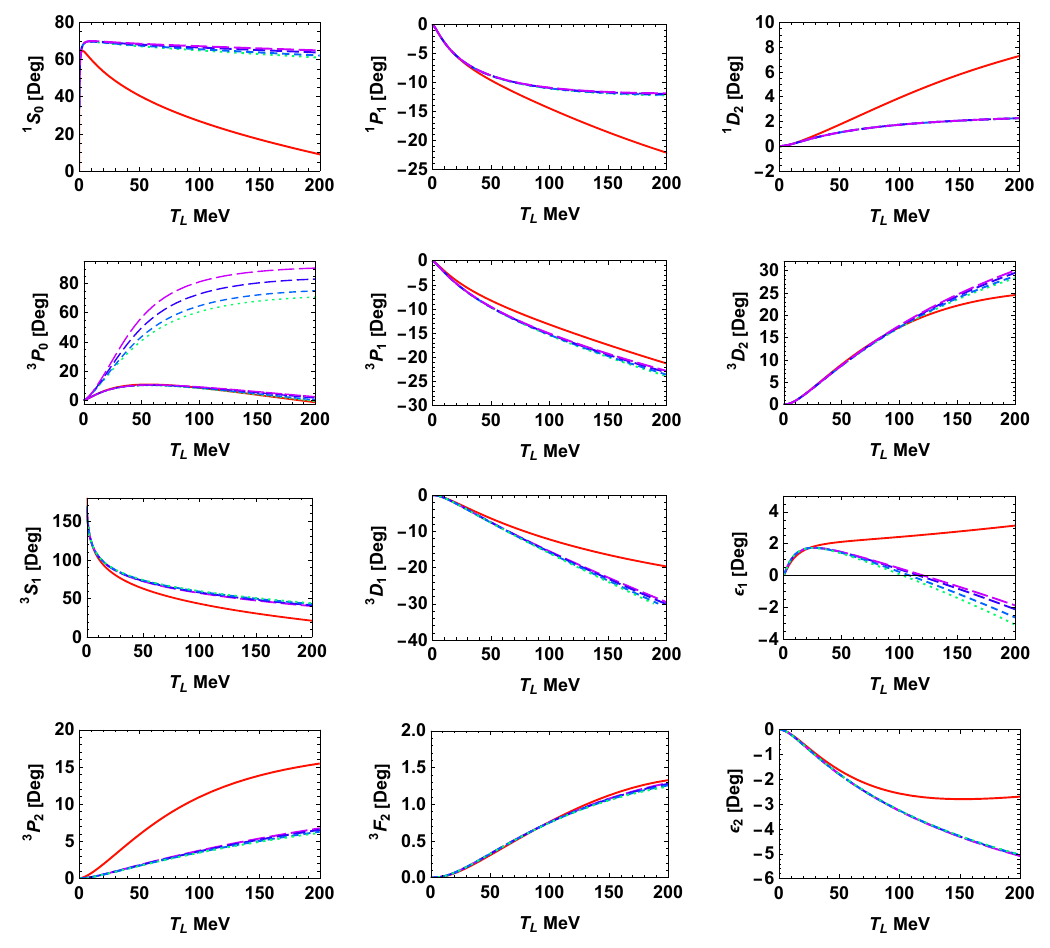}
\caption[]{\label{fig2} 
(Color online) Phase shifts and a mixing angle calculated at LO as functions of laboratory energy in
comparison with the Nijmegen \cite{Stoks:1993tb} partial wave analyses. Red solid lines correspond to the Nijmegen phase shifts and dashed lines,  (with growing lengths of dashes) represent our results for $\mu=350, \, 400, \, 500, \, 600$ MeV. In $^3P_0$ partial wave, we show two sets of our generated lines. The set close to the Nijmegen phase shifts represents our result calculated by promoting the leading contact interaction term to LO potential.}
\end{figure}

The obtained results are plotted in Fig.~\ref{fig2} together with  the phase shifts of the Nijmegen partial wave analyses \cite{Stoks:1993tb}. While our curves show mild smooth dependence on $\mu$, the description of the data is as expected from the leading order chiral effective field theory. In particular, the $^1S_0$ partial wave requires the promotion of the next-to-leading order contact interaction term to the leading order to achieve a more reasonable description of the empirical phase shifts at LO \cite{Epelbaum:2015sha}. We have not done so as it requires taking $n_N=2$, which is beyond the scope of the current work. On the other hand, as seen from the plots for the $^3P_0$ partial wave, promoting the next-to-leading order contact interaction improves the description of the data very significantly.\footnote{Promoting the next-to-leading order contact interaction term of the $^3P_0$ partial wave does not require taking $n_N=2$, as it is linear in both incoming and outgoing momenta, unlike the $^1S_0$ partial wave, where the next-to-leading order contact interaction is quadratic in both momenta.} 
We observe a mild cutoff dependence of the $^3P_2$ 
phase shifts compared to those obtained from typical potential-level regularizations~\cite{Nogga:2005hy,Epelbaum:2006pt}. This behavior stems from the higher-derivative regularization employed here, which leaves the Born term of the amplitude unmodified. Corrections from iterations of the LO potential are suppressed owing to its small magnitude. We also calculated the binding energy of the deuteron and obtained $ B_d =1.793, \, 1.808, \, 1.831, \, 1.846$ MeV for $\mu=350, \, 400, \, 500, \, 600$ MeV, respectively.

\section{The deuteron equation}
\label{deuteron_equation}

Below, we briefly summarize and modify accordingly the derivation of the equation for the
deuteron structure functions of Refs.~\cite{Epelbaum:2013naa,Panteleeva:2024abz}, to make it applicable to our higher-derivative formulation.  

We use the following definition of the deuteron interpolating field \cite{Kaplan:1998sz}:
\begin{equation}
{\cal D}_i\equiv N^T \mathcal{P}_i N=\sum_{\alpha,\beta,a,b=1}^2 N_{\alpha,a} \mathcal{P}_{i,ab}^{\alpha\beta} N_{\beta,b}
, \ \ \ \mathcal{P}_i\equiv \frac{1}{\sqrt{8}}\,{\sigma_2\sigma_i \tau_2},
\label{interfield}
\end{equation}
where $\alpha, \,\beta$ and $a, \,b$ are the spin and isospin indices, respectively.\footnote{Notice that observable quantities do not depend on a particular choice of interpolating fields.}
We are interested in the amplitude of the deuteron
interpolating field interacting with a pair of nucleon fields. This
quantity in the rest frame of the deuteron is given by 
\begin{equation}
D_j^{IJ}(\vec p\,') = \mathcal{P}_j^{IJ}  + \sum_{A,B=1}^{n_N+1}  \!\int\! d^3\vec k \  T^{IJ,AB} (\vec p\,'  , \vec k ) \, G^{AB}(\vec k) \, \mathcal{P}_j ^{AB}\,,
\label{TD}
\end{equation} 
where $\mathcal{P}_j ^{AB} = s_{A} s_B \mathcal{P}_j $, with $A,B \in \{1, 2\}$, and the amplitude $T ^{AB,IJ} $ is obtained by solving Eq.~(\ref{CChEqs}). 
It is easily seen that Eq.~(\ref{TD}) is equivalent to 
\begin{equation}
D_j^{IJ}(\vec p\,') = \mathcal{P}_j^{IJ}  + \sum_{A,B=1}^{n_N+1}  \!\int\! d^3\vec k \,  V^{IJ,AB} ( \vec p\,' , \vec k  ) \, G^{AB}(\vec k) \, D_j ^{AB}(\vec k)  \,.
\label{Dequation}
\end{equation} 
We parameterize the amplitude $D^{IJ} _j \left( \vec p\,'\right)$ in terms of two
structure functions $\Delta_1^{IJ} $ and $\Delta_2^{IJ} $ via 
\begin{equation}
D_{j}^{IJ} \left( \vec p\,'\right) = \Delta_1^{IJ} (\vec p\,'^2) \mathcal{P}_j^{IJ}  + 
 p'_a p'_b\, \Delta_2^{IJ} (\vec p\,'^2)\, \sigma^a \mathcal{P}_j^{IJ}  \left( \sigma^b \right)^T\, , 
\label{Dpar}
\end{equation}  
where we omit the isospin indices and do not show terms resulting from anti-symmetrization. 
To obtain equations for the structure functions $\Delta_{1,2}^{IJ} $, we parameterize the potential $V^{KL,IJ}$ by
indicating explicitly the spin indices: 
\begin{eqnarray}
V^{KL,IJ}_{\alpha\beta,\, \gamma\delta} \left(\vec p\,' ,\vec p\,\right)
& = & v^{0,KL,IJ}
\left( \vec p\,' ,\vec p \,\right)
\delta_{\alpha\gamma}\delta_{\beta\delta}+v^{1,KL,IJ}_a \left( \vec p\,' ,\vec p\,
\right) \,\left(\sigma^a_{\alpha\gamma}\delta_{\beta\delta} +
\delta_{\alpha\gamma}\sigma^a_{\beta\delta}\right)\nonumber\\
&+&v^{2,KL,IJ}_{ab} \left( {\vec p\,'} ,\vec p\,\right)
\,\sigma^a_{\alpha\gamma}\sigma^b_{\beta\delta}\,, 
\label{tpar}
\end{eqnarray}
where
\begin{eqnarray}
v^{0,KL,IJ}(\vec p\,',\vec p\,) & = & \nu_1^{KL,IJ}(\vec p\,',\vec p\,)\,,\nonumber\\
v^{1,KL,IJ}_a(\vec p\,',\vec p\,)  &=&  i\,\epsilon^{abc} p^b p'^c
\nu_3^{KL,IJ}(\vec p\,',\vec p\,)\,, \\
v^{2,KL,IJ}_{ab}(\vec p\,',\vec p\,) & = & \delta_{ab}\, \nu_2^{KL,IJ}(\vec p\,',\vec p\,) + p'^a p'^b\,
\nu_5^{KL,IJ}(\vec p\,',\vec p\,)+ p^a
p^b\, \nu_6^{KL,IJ}(\vec p\,',\vec p\,)  \nonumber\\
&+&(p^a p'^b+ p'^a p^b)\,\nu_4^{KL,IJ}(\vec p\,',\vec p\,),  \nonumber
\label{fversusampl}
\end{eqnarray}
and $\nu_i^{KL,IJ}(\vec p\,',\vec p\,)$ are scalar functions of ${\vec
p\,'}^2$, $\vec p\,^2$ and $\vec p\,'\cdot\vec p$. 
For the structure functions, we obtain the following system of integral equations: 
\begin{eqnarray}
\Delta_1^{IJ} (\vec p\,^2) & = & 1 + \sum_{K,L=1}^{n_N+1} \frac{ s_K s_L}{s_I s_J} \!\int\! d^3\vec k\, G^{KL}(\vec k)
\biggl\{  \Delta_1^{KL}(\vec k^2) \left[\nu_1^{IJ,KL}(\vec p,\vec k)+\nu_2^{IJ,KL}(\vec p,\vec k)+C_1 \nu_6^{IJ,KL}(\vec p,\vec k)\right]    \nonumber  \\
&+& \Delta_2^{KL}(\vec k^2) \biggl[  C_1\left(\nu_1^{IJ,KL}(\vec p,\vec k)+\nu_2^{IJ,KL}(\vec p,\vec k)+2 (\vec p\cdot \vec k) \nu_3^{IJ,KL}(\vec p,\vec k) \right)  \nonumber       \\
&+& 2\, \vec k^2 (\vec p\cdot \vec k)\, \nu_4^{IJ,KL}(\vec p,\vec k) \!+\! \left[(\vec p\cdot \vec k)^2-C_1 \vec p\,^2\right]\nu_5^{IJ,KL}(\vec p,\vec k)\!+\!(\vec k^2)^2 \nu_6^{IJ,KL}(\vec p,\vec k)\!\biggr]\!\biggr\},    \nonumber  \\
\Delta_2^{IJ} (\vec p\,^2) & = &  \sum_{K,L=1}^{n_N+1} \frac{s_K s_L}{s_I s_J}  \int d^3\vec k\, G^{KL}(\vec k)\,
\biggl\{\Delta_1^{KL}(\vec k^2) [2 B\, \nu_4^{IJ,KL}(\vec p,\vec k)+C_2 \nu_6^{IJ,KL}(\vec p,\vec k)+\nu_5^{IJ,KL}(\vec p,\vec k)]    \nonumber       \\
&+& \Delta_2^{KL}(\vec k^2)
\biggl[C_2 \left(\nu_1^{IJ,KL}(\vec p,\vec k)
+ \nu_2^{IJ,KL}(\vec p,\vec k)+2\, (\vec p\cdot \vec k)\, \nu_3^{IJ,KL}(\vec p,\vec k)\right) \nonumber\\
 &+& C_1 \nu_5^{IJ,KL}(\vec p,\vec k)
-2 B\, \vec k^2 \nu_3^{IJ,KL}(\vec p,\vec k)
\biggr]\biggr\},
\label{dsystem2}
\end{eqnarray}  
where 
\begin{equation}
B  \equiv  \frac{\vec p\cdot \vec k}{\vec p\,^2},\quad
C_1 \equiv  \frac{1}{2}\,\left[\vec k^2- \frac{(\vec p\cdot \vec k)^2}{\vec p\,^2}\right],\quad
C_2  \equiv  \frac{3\,(\vec p\cdot \vec k)^2-\vec k^2 \vec p\,^2}{2\,\vec p\,^4}.
\label{BCDef}
\end{equation}

\medskip

For the LO potential of Eq.~(\ref{OPETOPTNSt}) we have
\begin{eqnarray}
\nu_1^{KL,IJ}(\vec p\,',\vec p\,) & = & s_I s_J s_K s_L  \, C_S \,, \nonumber \\
\nu_2^{KL,IJ}(\vec p\,',\vec p\,) & = & s_I s_J s_K s_L  \, C_T  \,, \nonumber \\
\nu_3^{KL,IJ}(\vec p\,',\vec p\,) & = & 0\,, \nonumber \\
\nu_4^{KL,IJ}(\vec p\,',\vec p\,) & = & - s_I s_J s_K s_L \, \frac{g_A^2  }{4 F^2}\, \frac{\vec \tau_1\cdot \vec \tau_2}{2\,  \omega({\vec p\, '-\vec p}\,)} 
\nonumber\\ &\times& 
\left( \frac{1}{2 E_1(q) - E_I(p)-E_L(p') - \omega({\vec p\, '-\vec p}\,)}+\frac{1}{2 E_1(q) - E_J(p)-E_K(p') - 
\omega({\vec p\, '-\vec p}\,)}\right)   \,, \nonumber \\
\nu_5^{KL,IJ}(\vec p\,',\vec p\,) & = & s_I s_J s_K s_L \, \frac{g_A^2  }{4 F^2}\, \frac{\vec \tau_1\cdot \vec \tau_2}{2\, \omega({\vec p\, '-\vec p}\,)} 
\nonumber\\ &\times& 
\left( \frac{1}{2 E_1(q) - E_I(p)-E_L(p') -  \omega({\vec p\, '-\vec p}\,)}+\frac{1}{2 E_1(q) - E_J(p)-E_K(p') - 
  \omega({\vec p\, '-\vec p}\,)}\right)  \,, \nonumber \\
\nu_6^{KL,IJ}(\vec p\,',\vec p\,) & = & s_I s_J s_K s_L \, \frac{g_A^2  }{4 F^2}\, \frac{\vec \tau_1\cdot \vec \tau_2}{2\, \omega({\vec p\, '-\vec p}\,)} 
\nonumber\\ &\times& 
\left( \frac{1}{2 E_1(q) - E_I(p)-E_L(p') - \omega({\vec p\, '-\vec p}\,) }+\frac{1}{2 E_1(q) - E_J(p)-E_K(p') - \omega({\vec p\, '-\vec p}\,)}\right)  \,. 
\label{NLOCVpar}
\end{eqnarray} 
By solving the integral equations (\ref{dsystem2}) for the potential of Eq.~(\ref{NLOCVpar}) numerically, the binding energy and the structure functions of the deuteron can be obtained. By doing so we have reproduced the deuteron binding energy for various values of $\mu$ in good agreement with those specified in the previous section.

\section{Summary}
\label{sum}

To summarize, in this work we introduced a symmetry-preserving higher-derivative formulation of the non-relativistic baryon chiral perturbation theory. In the standard approach one writes down the most general Lagrangian consistent with underlying symmetries and then reduces the Lagrangian to its minimal canonical form by using various identities and field redefinitions. In our formulation, we keep selectively chosen terms with higher-order derivatives, which one could also get rid of by using field-redefinitions. We use these terms to obtain propagators with a modified ultraviolet behavior and thus tame ultraviolet divergences in our order-by-order calculations of physical quantities.  
While the choice (of an infinite number) of the higher-derivative terms, as well as the form of the canonical effective Lagrangian is fixed, for increasing chiral orders of the accuracy of calculations we take into account the higher-derivative terms with increasing number of derivatives. 
Thus, we achieve that at each chiral order, the integral equations of the few-body sector are free of ultraviolet divergences, provided that the corresponding effective potentials/kernels have been already renormalized. To regularize the remaining ultraviolet divergences appearing in zero and one-nucleon sectors as well as in diagrams contributing to effective potentials/kernels, we apply dimensional regularization. It follows by construction that the dependence of the full expressions of physical quantities on the parameters (coefficients) of the higher-derivative terms can be completely absorbed in the renormalization of the parameters of the Lagrangian. On the other hand, as in any quantum field theory, order-by-order perturbative results depend on the choice of the renormalization condition. In our case, this dependence is encoded in the dependence on the parameters of higher-derivative terms.

To obtain systematic diagrammatic rules, we present the propagators with modified ultraviolet behavior as linear combinations of ordinary propagators corresponding to real and fictitious particles with various masses. Substituting these linear combinations in Feynman diagrams and expanding term-by-term, we obtain Feynman diagrams for those fictitious particles with corresponding Feynman rules. By integrating Feynman diagrams over zeroth components of loop momenta, diagrams of the time-ordered perturbation theory (and thus the corresponding diagrammatic rules) are obtained. 

We have explicitly written down the coupled channel integral equations satisfied by the off-shell scattering amplitudes for two-nucleon sectors. 
While for on-shell amplitudes, the fictitious degrees of freedom never appear in external legs of diagrams, to obtain physical amplitudes one needs to solve coupled channel equations for off-shell amplitudes. This complicates numerical solution of the integral equations a lot, however does not make it unfeasible.       
As a proof of principle, we have calculated NN scattering phase shifts at leading order.  
We also obtained a system of integral equations satisfied by the structure functions of the deuteron. 

\acknowledgments
This work was supported in part by the Shandong Provincial Natural Science Fund for Excellent Young Scientists Fund Program (Overseas) with project no. 2025HWYQ-015, by the Key R\&D Program of Shandong Province (Grant No. 2026CXPT175), and by Qilu Youth Scholars Program of Shandong University, by the MKW NRW under the funding code NW21-024-A, by the Georgian Shota Rustaveli National Science Foundation (Grant No. FR-23-856), by the European Research Council (ERC) under the European Union’s Horizon 2020 research and innovation programme (grant agreement No. 885150) and by BMBF through the ErUM-Data project DEMOS.  The work of A.D. was supported  by the Alexander von Humboldt Foundation 
under Grant No. LTU-1185721-HFST-E.


\end{document}